\documentclass[a4paper, 10pt]{article}

\usepackage{graphics}
\usepackage{epsfig,amssymb,euscript,xspace, color}
\usepackage{amsmath,jheppub_0,mathtools,float}
\usepackage{romannum}
\def\bea{\begin{eqnarray}}
\def\eea{\end{eqnarray}}
\def\be{\begin{equation}}
\def\ee{\end{equation}}

\newcommand{\bra}{\langle}
\newcommand{\ket}{\rangle}
\newcommand{\rpl}{{r_{\!_+}}}
\newcommand{\rmi}{{r_{\!_-}}}
\newcommand{\Ll}{\mathcal{L}}



\begin{document}

\title{Fast Scrambling of Mutual Information in Kerr-$AdS_4$ }
\author{Vinay Malvimat$^a$, Rohan R. Poojary$^b$}
\affiliation{$^a$Theory Division, Saha Institute of Nuclear Physics, \\
Homi Bhaba National Institute (HBNI), \\
1/AF, Bidhannagar, Kolkata 700064, India. 
\vspace{0.3cm}
\\
$^b$Institute for Theoretical Physics, TU Wien,\\
Wiedner Hauptstrasse 8-10, 1040 Vienna, Austria.  }
\emailAdd{vinaymmp@gmail.com, rpglaznos@gmail.com}

\abstract{ We compute the disruption of mutual information between the hemispherical subsystems on the left and right CFT$s$ of a Thermofield Double state described by a Kerr geometry in $AdS_4$ due to shockwaves along the equatorial plane. The shockwaves and the subsystems considered respect the axi-symmetry of the geometry. At late times the disruption of the mutual information is given by the lengthening of the HRT surface connecting the two subsystems, we compute the minimum value of the Lyapunov index-$\lambda_L^{(min)}$ at late times and find that it is bounded by $\kappa=\frac{2\pi T_H}{(1-\mu\, \Ll)}$ where $\mu$ is the horizon velocity and $\Ll$ is the angular momentum per unit energy of the shockwave. At very late times we find the the scrambling time for such a system is governed by $\kappa$ with $\kappa t_*=\log \mathcal{S}$ for large black holes with large entropy $\mathcal{S}$. We also find a term that increases the scrambling time by $\log(1-\mu\,\Ll)^{-1}$ but which does not scale with the entropy of the Kerr geometry. }

\maketitle



\section{Introduction}
Chaos is a hallmark of large-$N$ many body dynamics. In recent years, the pursuit for understanding the mechanisms of chaotic phenomena in holographic systems has led to remarkable progress in quantum gravity. The initial works of Hayden and Preskill \cite{Hayden:2007cs} and Sekino \& Susskind  \cite{Sekino:2008he} had conjectured that black holes are amongst the fastest scramblers of in-fallen information based on quantum information theoretic arguments for systems with a large number of local degrees of freedom. Such systems were known to exhibit a scrambling time proportional to the local degrees of freedom $i.e.$
\be
t_*\sim\log \mathcal{S}, \hspace{0.4cm}{\rm where}\,\,\,\mathcal{S}={\rm entropy.}
\ee
A useful diagnostic of chaos is the Lyapunov index $\lambda_L$ which can be discerned from the late time decay of the out of time ordered correlators $OTOC$
\be
\frac{\bra W(0)V(t)W(0)V(t)\ket}{\bra WW\ket\bra VV\ket}=1-\epsilon \,e^{\lambda_L t}+\dots
\ee
where $\epsilon$ is a perturbative parameter in the theory. It can also be understood in terms of the scrambling of Mutual Information $I[A:B]=S_A+S_B-S_{A\cup B}$ between large enough subsystems in a Thermofield-Double (TFD) state describing an eternal black hole in AdS due to very early in-fallen perturbation. Here $S_A$ is the  entanglement entropy of a sub-region $A$ and  the two sub-regions $A$ \& $B$ are in the left and right CFTs describing the Hilbert space of the black hole in $AdS$. Mutual information can be regarded as a better diagnostic of chaos as it provides a bound for the correlation function of bounded operators \cite{Wolf:2007tdq}. 
For static black holes in 3d $AdS$ both these measures of chaos revealed a Lyapunov index given by their temperature $i.e.$ $\lambda_L=2\pi T_H$ \cite{Shenker:2013pqa}\cite{Shenker:2013yza}\cite{Shenker:2014cwa}. It was further  famously shown by Maldacena, Shenker and Stanford (MSS) \cite{Maldacena:2015waa} that large $N$ QFTs characterised by only temperature possessed an upper bound for chaos where $\lambda_L$ is to bounded by 
\be
\lambda_L\leq\frac{2\pi}{\beta},\hspace{0.5cm}{\rm where}\,\,\beta=T_H^{-1}.
\ee
thus validating the earlier conjecture of black holes as the fastest scramblers. The case for rotating black holes is more intriguing. 
Rotating black holes in $AdS_3$ were shown to posses 2 Lyapunov indices governing the behaviour of the OTOC of 4pt functions, each corresponding to the left and right moving temperatures of the dual CFT$_2$  \cite{Poojary:2018eszz}\cite{Jahnke:2019gxr}\cite{Stikonas:2018ane}. One of these temperatures is greater than the temperature of the BTZ and survives the extremality limit. The arguments made for arriving at the MMS bound were also generalized for large$-N$ thermal QFTs with a chemical potential $\mu$ associated with a global charge \cite{Halder:2019ric}, this revealed that  the Lyapunov index to be bounded by
\be
\lambda_L\leq\frac{2\pi/\beta}{(1-\mu/\mu_c)}
\label{Halder_bound}
\ee
where $\mu_c$ is the maximum or critical value attainable by $\mu$. It is also worth noting that such chaotic phenomena have been seen in other holographic systems such as the dynamics a string suspended in an AdS black hole \cite{deBoer:2017xdk}\cite{Banerjee:2018kwy}\cite{Banerjee:2018twd}. For the case of rotating black holes the string worldsheet degrees of freedom also see a chaotic behaviour consistent with the above bound \cite{Banerjee:2019vff}. 
However, for the case of rotating BTZ the long time behaviour of the OTOC seems to decay with an average value of $|\lambda_L|=2\pi T_H$ \cite{Mezei:2019dfv, Craps:2020ahu}. Thus the scrambling time for OTOCs in rotating BTZ and its holographic CFT$_2$ seems to exhibit the same scrambling time as in the static case with short bursts of scrambling where instantaneous $\lambda_L$ is greater of the 2 CFT temperatures.
However, the recent analysis of the scrambling of mutual information in rotating BTZ perturbed by shockwaves with non-zero angular momentum shows a different picture. Here rotating BTZ does exhibit a $\lambda_L>2\pi T_H$ even for late times and the scrambling time is also determined by such a $\lambda_L$ \cite{Malvimat:2021itk}.
\\\\
Yet another diagnostic of chaos in maximally chaotic systems is the 
 phenomena of ``pole skipping". Here the retarded energy 2pt functions in frequency space are seen to have poles except at points where the  frequency is $\omega=i\lambda_L$ and wave-number is $k=\pm i\lambda_L/ v_B$; here $v_B$ is the butterfly velocity which measures the spread of perturbation in space \cite{Grozdanov:2017ajz}\cite{Blake:2017ris}. Static black branes in $AdS$ have been shown to posses this phenomena of pole-skipping with $\lambda_L=2\pi T_H$.  This peculiar phenomena of pole-skipping has also been justified as arising from an effective theory of hydrodynamics of maximally chaotic systems \cite{Blake:2021hjj,Blake:2017ris}. It was also shown to be a generic feature of planar holographic black branes in $AdS$ \cite{Blake:2018leo}. For rotating BTZ the pole-skipping phenomena was studied in \cite{Liu:2020yaf} where the 2 Lyapunov indices corresponding to the 2 temperatures of the dual CFT$_2$ were found as the pole skipping points, corroborating the earlier results in rotating BTZ. Recent analysis of pole-skipping in Kerr-AdS$_4$  have uncovered $\lambda_L=2\pi T_H$ \cite{Blake:2021wqj}. 
\\\\
In this article  we analyse the effect of rotating shockwaves on the Mutual Information $I(A:B)$ between the left and right CFTs describing the analytically extended rotating bulk geometry as a thermofield double state. Here $A$ and $B$ are identical hemispheres which end at the equator at the left and right  boundary CFTs respectively. For simplicity the shockwaves considered are axi-symmetric $i.e.$ they exist for every value of the symmetric direction $\phi$, and are confined to the equator. The shockwave geometry to be solved for is given by the Dray-'tHooft solution provided one utilizes the Kruskal coordinates along the shockwave trajectory. We therefore construct such coordinates for  Kerr-$AdS_4$ metric along in-out going null rotating geodesics with angular momentum per unit energy $\mathcal{L}$, and write the Dray-'t Hooft solution at very late times as a function of the time $t_0$ in the past- denoting the time when the shockwaves were released. 
\\\\
Following the general construction by Shenker and Stanford \cite{Shenker:2013pqa} we then  analyse the late time behaviour of $I(A:B)$ due to such a shockwave perturbing the Kruskal extension via the left boundary at some time $t_0$ in the past $w.r.t$ the right CFT. Along the way we note that the blueshift suffered by such a rotating shockwave as it falls towards the horizon is given by $E\sim E_0e^{\kappa t_0}$ with 
\bea
\kappa=\frac{2\pi}{\beta(1-\mu\mathcal{L})},\hspace{0.3cm} {\rm where} &&{\mu=\Omega_+}={\rm horizon \,velocity,}\cr
&&\mathcal{L}={\rm angular\, mom./\, energy.}
\eea
This quantity is obtained by demanding that the Kruskal coordinates be smooth as one approaches the horizon and are devoid of any coordinate singularities. As observed in \cite{Shenker:2013pqa} the blueshift essentially dictates the late time behaviour of the perturbation to $I[A:B]$ and thus the scrambling of holographic mutual information between the 2 $CFT$s in a Thermofield double state describing the Kerr black hole in the bulk. The computation of $I[A:B]$  for this geometry is not analytically tractable and hence we determine the perturbation to $S_{A\cup B}$ \cite{Leichenauer:2014nxa} at very late times as it is the only term in $I(A:B)$ which is sensitive to the shift in horizon along null coordinates, unlike $S_A$ and $S_B$ which remain unaffected. We obtain this perturbation at $t=0$ as function of $t_0$ and  the black hole parameters, and find that it grows linearly in $t_0$ as expected $i.e.$
\be
I[A:B]\sim\mathcal{A} \,\lambda_L\, t_0 
\ee
where $\mathcal{A}$ can be taken to be the unperturbed area capturing $S_{A\cup B}$. We would regard the above $\lambda_L$ as the instantaneous Lyapunov index computed at late times $i.e.$ $\gg\beta=T_H^{-1}$. We find that for $\mathcal{L}=0$, $\lambda_L^{\rm (min)}\leq\kappa_0=2\pi T_H$ with the equality holding at extremality. For general values of $\mathcal{L}$ we find 
\be
\lambda_L^{\rm (min)}\leq\kappa=\frac{2\pi T_H}{(1-\mu\mathcal{L})}
\ee
We plot this as a function of the ratio of the inner to outer horizon $\rmi/\rpl$; Fig [\ref{F_cH_c}]. 
We also find that  as one approaches extremality, after  a finite value of $\rmi/\rpl$, $\lambda_L^{\rm (min)}>2\pi T_H$ $i.e.$ the black hole scrambles information at a rate greater than $2\pi$ times its temperature. 
The near extremal limit of $\kappa=\kappa_{ext}$ can be non-zero if $\mathcal{L}\rightarrow \mu^{-1}$ as one approaches extremality. For such a limit we find 
\be
0<\lambda_L^{\rm (min)}<\kappa_{ext}.
\ee 
We find that at very late times the rate of scrambling is given by $\kappa$ and the scrambling time by
\be
t_*=\frac{1}{\kappa}\log [\mathcal{S}]+\dots
\ee
where $\mathcal{S}$ is the entropy of the black hole and the $\dots$ indicate terms which do not scale with $\mathcal{S}$ at large values of entropy.
\\\\
This paper is organized as follows: In section 2 we review some details of the Kerr black hole in $AdS_4$. In section 3 we introduce the rotating null coordinates appropriate for solving the Dray-'tHooft solution for a rotating shockwave in this Kerr geometry. In doing so we compute the blueshift along a null in-falling geodesic with angular momentum $\mathcal{L}$. We also solve for the Dray-'t Hooft solution for an equatorial axi-symmetric shockwave with angular momentum $\mathcal{L}$. In section 4 we turn to compute the perturbation of the Mutual Information between the CFT$_L$ and CFT$_R$ for identical hemispheres ending at the equator due to such a shockwave.  We end with some Discussions and Conclusions in section 5.
\section{Kerr-$AdS_4$}
The Kerr metric in $AdS_4$ in the Boyer-Lindquist coordinates  \cite{Plebanski:1976gy,Carter:1968ks,Iyer:1994ys} takes the form
\bea
&&ds^2=\rho^2\left(\frac{dr^2}{\Delta}+\frac{d\theta^2}{\Delta_\theta}\right)-\frac{\Delta}{\rho^2}\left(dt-\frac{a\sin^2\theta}{\Xi}d\phi\right)^2+\frac{\Delta_\theta\sin^2\theta}{\rho^2}\left(adt-\frac{r^2+a^2}{\Xi}d\phi\right)^2\cr&&\cr
{\rm where}&&
\hspace{2cm}\Delta=(r^2+a^2)(1+r^2/l^2)-2mr,\hspace{0.5cm}\Xi=1-\frac{a^2}{l^2},\cr&&\cr
&&\hspace{2cm}\rho^2=r^2+a^2\cos^2\theta,
\hspace{0.5cm}\Delta_\theta=1-\frac{a^2}{l^2}\cos^2\theta.
\label{Kerr_Boyer_Lindquist}
\eea
The outer horizon exists as the largest root of $\Delta=0$.
Here the boundary coordinates at $r\rightarrow\infty$ are rotating with an angular velocity of
\be
\Omega_\infty=-\frac{a}{l^2}
\label{vel_boundary}
\ee
while the horizon rotates at
\be
\Omega_\phi=\frac{a\,\Xi}{\rpl^2+a^2}
\label{vel_horizon}
\ee
Therefore the horizon's velocity $w.r.t.$ a stationary boundary observer is
\be
\Omega_+=\Omega_{\phi}-\Omega_\infty=\frac{a(1+\rpl^2/l^2)}{\rpl^2+a^2}=\mu
\label{vel_actual}
\ee
which is indeed the chemical potential relevant for the thermodynamics of the black hole \cite{Papadimitriou:2005ii}. The temperature and entropy can be given in terms of $\{a,\rpl\}$ as
\be
2\pi T_H=\frac{\rpl^2-a^2+\rpl^2/\l^{-2}(3\rpl^2+a^2)}{2\rpl(\rpl^2+a^2)},\hspace{0.2cm}\mathcal{S}=\frac{\pi(\rpl^2+a^2)}{G_N\Xi}
\ee
The rotating black hole in $AdS$ would correspond to a boundary large $N$ theory with the above fixed temperature and chemical potential given by $\beta \mu$ where $T_H=\beta^{-1}$. The bulk geometry provides a semi-classical description of a system with a partition function and density of states  given by
\be
Z=Tr\left[e^{-\beta H-\beta \mu J}\right],\hspace{0.4cm}\rho=\frac{e^{-\beta H-\beta \mu J}}{Z}
\ee
Following \cite{Maldacena:2001kr}, in holography black holes in $AdS$ can be purified by considering its Kruskal extension, in such a description the two-sided $AdS$ black hole is described by a pure state in the extended Hilbert space of the 2 boundary CFTs at the left and right exteriors
\be
|\Psi_{\rm BH}(\beta,\mu)\rangle=\frac{1}{\sqrt{{\rm Tr}\,e^{-\beta(H+\mu J)}}}\sum_i e^{-\beta(H+\mu J)}\,\,|i_L\rangle \otimes |i_R\rangle.
\ee 
The black hole entropy in any of the exteriors can therefore be obtained as the von Neuman entropy of the density of states after having integrated out one of the CFT Hilbert space. The dual bulk geometry describes a fine grained atypical correlation between the left and the right CFTs. This non-trivial entanglement between the 2 CFTs is characterized by the holographic entanglement entropy which at the semi-classical level is given by the area of the Ryu-Takayanagi (RT) surface for the case of static geometries and by the Hubeny-Rangamani-Takayanagi (HRT) surfaces for the case of time dependent geometries \cite{Ryu:2006bv,Hubeny:2007xt}. 
\\\\
A useful diagnostic of this fine correlation is the Mutual Information $I[A:B]$ between space-like subsystems $A$ in the CFT$_L$ and $B$ at the CFT$_R$ 
\be
I[A:B]=S_A+S_B-S_{A\cup B}
\ee  
where  $S_{A_i}$ in the above equation denotes is the entanglement entropy of the subsystem-$A_i$ obtained by tracing out the rest of the system from the density matrix. As both RT and HRT surfaces are minimal area co-dimension two surfaces homologous to the boundary subsystems, the extent they probe the bulk depends on the sizes of the subsystems. For small enough subsystems $A$ \& $B$ this implies that $I[A:B]$ vanishes as $S_{A}=S_{B}$. This is simply because the other possible minimal area surface homologous to $A\cup B$ traversing the black hole geometry would be larger than $S_A+S_B$. However for large enough  sizes of $A$ \& $B$ the situation reverses and $S_{A\cup B}$ is given by an HRT surface traversing the two boundaries. In such cases we get a glimpse of the fine correlation between the 2 subsystems especially when we observe the behaviour of $I[A:B]$  for late times after having perturbed it with an $\mathcal{O}(1)$ operator in the CFT or equivalently an $\mathcal{O}(G_N)$ perturbation in the bulk \cite{Shenker:2013pqa}. 
\\\\
The study of perturbations to black hole geometries can be very complicated and is a fertile area of research in numerical relativity, however certain well known exact solutions exist when the perturbations are shockwaves $i.e.$ the Vaidya geometry \cite{Vaidya:1950} and the Axelburg-Sexl or the Dray-'tHooft solutions \cite{Dray:1984ha,Sfetsos:1994xa}. A small perturbation emanating from the boundary of $AdS$ in a black hole geometry would get blue-shifted as time passes and it gravitates towards the horizon. Therefore at very late times $\gg\beta$, such a perturbation can be regarded as a shockwave and- as observed by \cite{Shenker:2013pqa}, can be utilized to compute the change in the late time behaviour of mutual information for large enough subsystems due to such a perturbation. As noted in \cite{Shenker:2013pqa} the late time change in $I[A:B]$ is basically governed by the blueshift suffered by the in-falling shockwave. For static geometries this is simply given by
\be
E\sim E_0e^{\frac{2\pi}{\beta} t_0},
\ee 
where $E$ is the energy of the shockwave as it passes the bifurcate horizon and $E_0$ and $t_0$ are the energy  at the boundary and the time in the past when the shockwave was released. We release this shockwave from the left boundary at time $t_L=t_0\gg\beta$ which can be regarded as perturbing right system in its far past. The mutual information at $t_L=0=t_R$ for large enough subsystems would see the effect of this shockwave due the the above blueshift and is scrambled at a rate determined by the index of the blue-shift $i.e.$ the temperature $\frac{2\pi}{\beta}$. The blueshift of the shockwave is tied in with the Kruskal coordinates used to analyse the problem. One generally chooses
\bea
&&U=e^{\frac{2\pi}{\beta}(r_*-t)},\hspace{0.4cm}V=e^{\frac{2\pi}{\beta}(r_*+t)}\hspace{0.5cm}
{\rm where}\,\,\,r_*\sim-\int_\infty^{r'}	\frac{dr}{g_{tt}}
\eea
is the tortoise coordinate. However, this decides the trajectory of the null shockwave as the Dray-'tHooft solutions are given for shockwaves at $U=0$ or $V=0$ $i.e.$ for shockwaves along the past or the future horizons \cite{Sfetsos:1994xa}.  For a more complete analysis, in the present article we consider the Kerr geometry perturbed by the rotating shockwaves with angular momentum per unit energy $\mathcal{L}$. The analysis of such a perturbation along the lines of Dray-'tHooft in this case would therefore require the Kruskal coordinates which also follow the shockwave.
\section{The Dray-'tHooft solution}
In this section we construct the Dray-'tHooft solution for a generic null shockwave as a function of the time $t_0$ when it was released from one of the boundaries. For simplicity we would be considering shockwaves along the equatorial plane $\theta=\pi/2$ such that they always stay on the equator. This is guaranteed by the symmetry of the geometry. We first begin by constructing the Kruskal coordinates along such null rotating geodesics with angular momentum $\Ll$ along the equator and then extend it to arbitrary values of $\theta$. In this process we uncover the index of the exponential blueshift suffered by the shockwave. We then construct the Dray-'tHooft solution by determining the shift in one of the Kruskal coordinates in terms of the initial time $t_0\gg\beta$ when it was released into the bulk.  
\subsection{Rotating Kruskal coordinates at the equator} 
In falling null  geodesics $\xi\cdot\partial$ would be parametrised by energy $\mathcal{E}=1$, angular momenta about the $\phi$ direction and the Carter's constant $\mathcal{Q}$ given by the Killing-Yano tensor $K_{\mu\nu}$
\be
\xi^2=0,\hspace{1cm}g_{\mu\nu}\xi^\mu\zeta^\nu_t=\mathcal{E},\hspace{1cm}g_{\mu\nu}\xi^\mu\zeta^\nu_\phi=\mathcal{L},\hspace{1cm}\xi^\mu K_{\mu\nu}\xi^\nu=\mathcal{Q}
\label{geodesic_cond}
\ee
where $\zeta_t=\partial_t-\tfrac{a}{l^2}\partial_\phi$ and $\zeta_\phi=\partial_\phi$. 
The geodesics along the equator stay along the equator. We define in-falling(out-going) Kruskal coordinates along such geodesics with arbitrary $\mathcal{L}$. It turns out that $\xi^\theta$ can consistently be put to zero at  the equator for 
\be
\mathcal{Q}=-\frac{(a-\mathcal{L})^2}{a^2}
\ee 
in \eqref{geodesic_cond}. The metric \eqref{Kerr_Boyer_Lindquist} can then be  recast using the one-forms dual to the in-out double null geodesics to look like
\bea
&&ds^2_{\theta=\tfrac{\pi}{2}}=F\,\xi_\mu^+\xi_\nu^- dx^\mu dx^\nu + h\,(dz+h_\tau d\tau)^2 \cr&&\cr
&&\tau=t\left(1-\tfrac{a\mathcal{L}}{l^2}\right)-\mathcal{L}\,\phi,\hspace{1cm} \eta z=(1+\mathcal{L}\gamma)\,\phi-(1-\tfrac{a\mathcal{L}}{l^2})\gamma \,t
\label{Kerr_LC_equator}
\eea
where the first term is obtained by comparing the coefficient of the  $dr^2$ term in \eqref{Kerr_Boyer_Lindquist}. Here the shift of the $z$ coordinate in $t$ is forced by demanding that $h_\tau\sim(r-\rpl)$ as $r\rightarrow\rpl$ which fixes $\gamma$ to be
\be
\gamma=\frac{a(1-a^2/l^2)}{(a^2+\rpl^2)l^2-(\rpl^2+l^2)a\mathcal{L}}
\ee
$\eta$ can be chosen to one, however choosing 
\bea
&&\eta=\frac{l^2(1+\mathcal{L}\gamma)}{l^2-a\mathcal{L}},\hspace{0.3cm}{\rm implies}\cr&&\cr
&&t=\frac{\tau+\Ll\, z}{1-\mu \,\mathcal{L}},\hspace{0.3cm}\phi=\frac{(1-a\mathcal{L}/\l^2)\,z+\Omega_{\phi}\tau}{1-\mu \,\mathcal{L}}.
\label{t_phi_to_tau_z}
\eea
The choice of $\eta$ or $z$ is such that we recover the horizon area when $\theta$ is integrated from $0$ to $\pi$ and $z$ from $0$ to $2\pi$ for the above metric at the outer horizon\footnote{The reason we choose the horizon and not any other point is because in calculating the Dray-'tHooft solution we would be requiring the transverse components be smooth across the horizon, also the shockwave would produce the strongest backreaction at $r=\rpl$ than at any other value of $r$ and it is this near horizon dynamics we wish to capture. Note, $z$ is precisely the near horizon co-moving coordinate one needs to utilize in order to work in the near horizon region of (near)extremal Kerr geometries $c.f.$ \cite{Lu:2008jk}\cite{Hartman:2008pb}.}.
The above relation can be inverted 
\be
\tau=\bigg(1-\frac{a\Ll}{l^2}\bigg)t-\Ll\, \phi, \hspace{0.4cm}z=\phi-\mu \,t.
\label{tau_z_to_t_phi}
\ee
This choice of $\eta$ 
also implies that $z$ is a co-moving co-ordinate along the horizon's angular velocity as observed by a stationary observer at the $AdS$ boundary. The co-moving coordinate $z$ also occurs when one takes the near horizon limit of extremal Kerr solutions \cite{Guica:2008mu,Lu:2008jk}.
The null vector fields $\xi_\pm$ have their signs of $\mathcal{E}$ \& $\mathcal{L}$ reversed $w.r,t.$ each other. The light-cone coordinates $\{u,v\}$ are given in terms of the one-forms
\bea
&&du=\xi^+_\mu dx^\mu \implies u=r_*-\tau\cr&&\cr
&&dv=\xi^-_\mu dx^\mu \implies v=r_*+\tau
\label{LC_coordinates}
\eea
We now note that the tangent vector $\chi=\partial_u$ (or $\partial_v$) is not affine at $\rpl$ $i.e.$ $\chi\cdot\nabla\chi^\mu=\mathcal{K}\chi^\mu$, where
\be
\mathcal{K}=\tfrac{1}{2}\xi^+\cdot\partial F
\ee
We can therefore define new coordinates $\{U,V\}$ on the equator such that they are affine at $\rpl$
\be
U=-e^{\kappa(r_*-\tau)},\,\,V=e^{\kappa(r_*+\tau)},\hspace{0.7cm}{\rm where} \,\,\kappa=\mathcal{K}\vert_{\rpl}
\label{Kruskal_coordinates_equator}
\ee
These are the required Kruskal coordinates with angular momenta $\mathcal{L}$. The above co-ordinates capture the right exterior while the left exterior is obtained by changing the signs in front of $\{U,V\}$. The metric now takes the form
\be
ds^2_{\theta=\tfrac{\pi}{2}}=\frac{F}{\kappa^2 UV}dUdV+ h\,(dz+h_\tau d\tau)^2 
\label{Kerr_Kruskal_equator}
\ee
The blueshift along this in falling  null geodesic is given by the exponential factor $e^{\kappa t_0}$ for some particle released from the boundary at time $t=0$.
We obtain the following result for
generic values of ${\cal L}$ 
\bea
\kappa=\frac{\kappa_0}{(1-\mu\mathcal{L})},\,\,\,\,{\rm where}\,\,\mu=\Omega_+
\label{kappa_0}
\eea
For the specific case of  a shockwave with zero angular momentum  $i.e$ ${\cal L}=0$ we get

\bea
&&\kappa\vert_{\mathcal{L}\rightarrow0}=\kappa_0=\frac{2\pi}{\beta}=2\pi T_H,\,\,\,\,
\eea

\subsubsection{Turning point}
\label{turning_point_analysis}
It must be noted that not any value of $\mathcal{L}$ is allowed as we will later setup the shockwave to emanate from the $AdS$  boundary and reach at least the outer horizon. Assuming that we send null perturbations from the very start\footnote{This is not required as one may choose to send in a massive particle who's trajectory tends to that of a null particle as it approaches the horizon.} the allowed values of $\mathcal{L}$ for such null geodesices are obtainable by a turning point analysis. We find that for $\mathcal{L}>0$ the following general conditions hold
\bea
\mathcal{L}&\leq&\Ll_{max}<\mu^{-1},\hspace{0.4cm}  {\rm for}\,\,T_H>0\cr&&\cr
\mathcal{L}&\leq&\Ll_{max}=\mu^{-1},\hspace{0.4cm}  {\rm for}\,\,T_H=0
\label{turning_point}
\eea  
where $\Ll_{max}$ is the maximum allowed limit of angular momenta (per unit energy) beyond which the turning point is outside the horizon.
The above condition at extremality can be checked explicitly by plugging in $\mathcal{L}=\mu^{-1}$ into the  radial component $\xi^r=\dot{r}$ of the null geodesic to find $\xi^r(\rpl)=0$.  
Therefore $\kappa$ doesn't blow up and is finite and non-zero even at extremality. We obtain the extremal value of $\kappa$ by scaling $\rmi\rightarrow \epsilon \rpl$, $\mathcal{L}\rightarrow \epsilon \mu^{-1}$ and taking the limit $\epsilon\rightarrow 1$
\be
\kappa_{\rm ext}=\frac{1+6r_0^2/l^2-3r_0^4/l^4}{4r_0(1+r_0^2/l^2)}
\label{extremal_blueshift}
\ee
where $r_0$ is the extremal horizon radius. This is akin to taking the extremal limit of $\Ll=\frac{\rmi}{\rpl}\mu^{-1}$. We find that 
\be
\Ll=\frac{\rmi}{\rpl}\mu^{-1}\leq \Ll_{max}
\ee
with the equality holding at extremality.
We  also observe that the extremal limit of $\kappa$ for the above form of $\Ll$ is related to the left moving Frolov-Thorne temperature $T_L$ observed in the Kerr-CFT literature for Kerr-$AdS_4$ at extremality \cite{Lu:2008jk} 
\be
\kappa_{\rm ext}=\frac{T_L}{2\,\Omega_{\phi\,{\rm ext}}}.
\label{ext_kappa}
\ee 
Here $\Omega_{\phi\,{\rm ext}}$ is the extremal value of the horizon velocity \eqref{vel_horizon} measured in the  rotating Boyer-Lindquist frame at the boundary. Note that $\kappa_{\rm ext}$ depends on how we scale $\Ll$ $w.r.t.$ the black hole parameters as we approach near extremal configuration, in general it can easily be seen that
for $\rmi\rightarrow \epsilon\rpl$ and $\Ll\rightarrow\epsilon^\sigma\mu^{-1}$ we find the above $\kappa_{\rm ext}$ changes to
\be
\kappa_{\rm ext}\rightarrow\frac{\kappa_{\rm ext}}{\sigma}
\ee
It would be interesting to understand how different possible ways of scaling $\Ll$ relate with regards to the near horizon perspective. 
\subsection{Full Coordinates}
We would like to extend the above set of coordinates along the entire sphere. Although this would not be required for the later computations we demonstrate here that this is indeed possible.
The above set of coordinates were written out for the equatorial plane at $\theta=\pi/2$, however in order to truly realize $\{U,V\}$ as Kruskal coordinates we need to show that they are exact $i.e.$
\be
\xi^{-}_\mu dx^\mu = dV,\,\,\, \xi^{+}_\mu dx^\mu = dU, \,\,\,\forall\, \theta\in [0,\pi]
\ee
For this it is enough to show
\be
\partial_\theta\xi^\pm_r=\partial_r\xi^\pm_\theta
\ee
This implies $\mathcal{Q}=const$. As we would be interested in the scenario wherein the shock-wave exists only on the equatorial plane we choose this to constant to be 
\be
\mathcal{Q}=-\frac{(a-\mathcal{L})^2}{a^2}
\ee 
as this condition is equivalent to choosing the $\xi^\theta\vert_{\pi/2}=0$. Note that this implies that the non-equatorial geodesics have $\xi^\theta\neq 0$ $i.e.$ they do not move along constant values of $\theta$.
It turns out that for constant value of $\mathcal{Q}$ above we have $\partial_\theta\xi^\pm_r=\partial_r\xi^\pm_\theta=0$.
We further have to have to reverse the sign of the $\theta$ component  between in and out going null geodesics $i.e.$ $\xi^\theta_+=-\xi^\theta_-$ as simply reversing the signs of $\mathcal{E}$ \& $\mathcal{L}$ doest seem to do this. This is important as we expect in going geodesics to have the opposite rate of change in $\theta$ as compared to the out going one.
\\\\
The metric then takes form
\bea
ds^2=\frac{F}{\kappa^2 UV}dUdV + h\,(dz+h_\tau d\tau)^2 +g\, (d\theta+g_\tau d\tau)^2,\hspace{0.4cm} g_\tau(\pi/2)=0 
\label{Kerr_Kruskal}
\eea
where
\bea
&&F(r,\theta)=\cr&&\cr
&&\hspace{0.2cm}=\frac{\ell ^2 \left(a^2 \left(r^2+\ell ^2\right)+r \ell ^2 (r-2 M)+r^4\right) \left(a^2 \cos ^2(\theta )+r^2\right)}{r \left(\ell ^4 \left(a^2 (2 M+r)+r^3\right)+\mathcal{L}^2 \left(a^2 r \left(r^2+\ell
   ^2\right)-\ell ^2 \left(\ell ^2 (r-2 M)+r^3\right)\right)-a^2 r \ell ^2 \left(a^2+r^2\right)-4 a M \ell ^4 \mathcal{L}\right)}\cr&&
\eea 
Here we have used the functions $\{F,h,h_\tau,g,g_\tau\}$ which now depend on $r$ and $\theta$. We use the same letters to denote these functions as in the previous section with the understanding the analysis in the previous section was carried out at $\theta=\pi/2$. The rest of the functions can be easily deduced from the transformation \eqref{t_phi_to_tau_z}.
The transverse directions can be worked out in detail and we do not explicitly write them here. We concern ourselves with the smoothness of the $\{U,V\}$ directions. It is straightforward to see that the Kruskal coordinates are defined as before  to be
\be
U=-e^{\kappa(r_*-\tau-\tilde{g}(\theta))},\,\,V=e^{\kappa(r_*+\tau+\tilde{g}(\theta))},\hspace{0.7cm}{\rm where} \,\,\kappa=\mathcal{K}\vert_{\rpl}
\label{Kruskal_coordinates_equator}
\ee
with an extra $\theta$ dependence
\be
\widetilde{g}(\theta)=\frac{\sqrt{-\Xi  \cot ^2 \theta  \left(\ell ^2 (\Xi -\Delta_\theta )+\Delta_\theta  \mathcal{L}^2\right)}}{\Delta_\theta }\implies \widetilde{g}(\pi/2)=0
\ee
$\mathcal{K}$ is similarly defined as
\be
\mathcal{K}=\tfrac{1}{2}\xi^+\cdot\partial F
\label{Kappa}
\ee
The extra $\theta$ dependence is absorbed into $F$ however $\kappa$ is independent of $\theta$ and is still given by
\be
\kappa=\frac{\kappa_0}{(1-\mu\mathcal{L})},\hspace{0.7cm}{\rm where}\,\,\,\, \kappa_0=\frac{2\pi}{\beta},\,\,\mu=\Omega_+
\label{kappa_1}
\ee
The above value of $\kappa$ suggests an upper bound for the angular momentum of the shock-wave to be $\mathcal{L}<\mu^{-1}=\Omega_+^{-1}$. 
\\\\
The functions $h_\tau$ and $g_\tau$ vanish at the future or the past outer horizon. The redefinition of $\phi\rightarrow z$ is precisely dictated by this condition. Note that $\theta$ doesn't need to be redefined. This requirement would be crucial for writing down the Dray-'t Hooft solutions in response to a shockwave.
\\\\
The specific forms of the functions $\{F,h,h_\tau,g,g_\tau\}$ are quite cumbersome, it is useful to remember their  behaviours close to the outer horizon
\be
F\sim F'(\rpl)(r-\rpl),\hspace{0.4cm} h_\tau\sim \mathcal{O}(r-\rpl),\hspace{0.4cm} g_\tau\sim\mathcal{O}(r-\rpl)
\ee
while the rest approach constant values.
We note the values of $F'(\rpl)$ below 
\bea
  &&F'(\rpl)=\frac{\pi T_H\, (\rpl^2+a^2\cos^2\,\theta)}{(a^2+\rpl^2)(1-\mu\,\Ll)^2}\hspace{0.4cm} 
\eea
It would be necessary for later purposes to expand $F(r,\theta)$ around $\rpl$ and re-write it in terms of $UV$. Expanding $F$ as
\be
F(r,\theta)=(r-\rpl)F'(\rpl)+\tfrac{1}{2}(r-\rpl)^2F''(\rpl)+\dots
\ee
we choose
\be
(r-\rpl)F'(\rpl)=-\mathbb{A}\,\, UV.
\ee
The $-$ve sign is chosen as $lhs$ above is positive and $UV$ is negative in the exterior regions.
The proportionality constant $\mathbb{A}$ is an unknown function of the black hole parameters who's precise form is not required for the analysis, however we do require it not scale $w.r.t.$ the entropy of the black hole. This is expected as the above equation is simply a relation between coordinates. 
 Therefore we find
\be
\frac{F}{\mathbb{A}\, UV}=-1-\frac{\mathbb{A} \,UV}{2} \frac{F''(\rpl)}{F'(\rpl)^2}+\dots
\ee
where $\dots$ denote $\mathcal{O}(U^2V^2)$ terms.
\subsubsection{Kerr-Newman $AdS_4$}
The generalization to the charged case is straightforward. The metric is the same as in \eqref{Kerr_Boyer_Lindquist} with the exception of 
\be
\Delta=(r^2+a^2)(1+r^2/l^2)-2mr +q^2
\ee 
and a gauge field given by
\be
A=-\frac{2qr}{\rho^2}\left(dt-\frac{a \sin^2\theta}{\Xi}d\phi\right).
\ee
The electric charge is given by
\be
Q=\frac{2q}{\Xi}
\ee
The expressions for the angular velocities is the same as in the Kerr case in terms of $a$ and $\rpl$. 
The blueshift along a null in-falling geodesic is then  given by
\be
\kappa=\frac{2\pi T_H}{(1-\mu\mathcal{L})},\hspace{0.2cm}{\rm where}\,\,\,\,2\pi T_H=\frac{\rpl^2-(a^2+q^2)+\rpl^2/\l^{-2}(3\rpl^2+a^2)}{2\rpl(\rpl^2+a^2)}.
\ee
It is interesting to note that the form of the blueshift is unchanged and is only dependent on the coupling $\mu\,\mathcal{L}$ and $T_H$. Note, the trajectory of charged and un-charged in-falling massive particles at sufficiently late times would still be that of a null geodesic. However, the amount of time required for a charged massive  particle to approach a light-like trajectory would be different from that of an uncharged one. Massless particles on the other hand would follow light-like trajectories irrespective of their charge.  We also repeat similar analysis for the case of RN $AdS_4$ in the appendix \ref{RN} and find that the angular momentum of the shockwave has no effect on the blueshift seen at the horizon.  
In the rest of the paper we only consider the Kerr $AdS_4$ case for simplicity as the analysis can be readily generalised.
\subsection{Shockwaves along the equator}
We next consider in-falling shockwaves along the equatorial plane with angular momentum $\mathcal{L}$. The coordinate system developed in the previous subsection is precisely suited for studying the back reaction generated by these shockwaves. Like in the case studied in \cite{Shenker:2013pqa} we choose some time $t_0$ in the far future of the left exterior at which we send in an axisymmetric shockwave. We begin with the Dray-'t Hooft solution for the metric \eqref{Kerr_Kruskal} for a axi-symmetric shockwave at $U=0$ $i.e.$ along the $V$ coordinate which is in-falling\footnote{It is regarded in-falling $w.r.t.$ the time in the right exterior.} in the left exterior. The metric for $U<0$ is defined by \eqref{Kerr_Kruskal} while the metric for $U>0$ is given by shifting 
\be 
V\rightarrow \widetilde{V}=V+\alpha\,\Theta (U) f(\theta)
\label{V_shift_0}
\ee
where the constant $\alpha$ is proportional to the strength of the shockwave at $\theta=\pi/2$; its precise form will be  obtained by satisfying a smoothness condition. The response function $f(\theta)$ captures the response of the shockwave in the transverse direction and is independent of $\phi$ due to axi-symmetry. The metric \eqref{Kerr_Kruskal} in the coordinates $\{U,V,z\}$ can be obtained by noting
\be
d\tau=\frac{1}{2\kappa UV}(UdV-VdU)-\widetilde{g}'d\theta
\ee  
yielding
\bea 
ds^2&=&\frac{F}{\kappa^2 UV}dUdV+ h\,\left[dz+\frac{h_\tau}{2\kappa UV}(UdV-VdU)-h_\tau\widetilde{g}'d\theta\right]^2 +\cr&&\cr
&&\hspace{2.8cm}+g\, \left[d\theta+\frac{g_\tau}{2\kappa UV}(UdV-VdU)-g_\tau\widetilde{g}'d\theta\right]^2
\eea
The Dray-'tHooft solution is then simply obtained by implementing the shift \eqref{V_shift_0}
\be
ds^2\rightarrow \widetilde{ds}^2 - \frac{F}{\kappa^2 UV}\delta(U)f dU^2
\ee
where $\widetilde{ds}$ denotes the line element \eqref{Kerr_Kruskal} with $V\rightarrow\widetilde{V}$  \cite{Sfetsos:1994xa}. 
The above metric solves the Einstein's equation in presence of a stress-tensor sourced by a shockwave
\be
R_{\mu\nu}-\frac{1}{2}Rg_{\mu\nu}-\Lambda g_{\mu\nu}=R_{\mu\nu}+\frac{3}{l^2}g_{\mu\nu}=-8\pi G_N T_{\mu\nu}
\ee
\be
\alpha\mathcal{D}_{\theta}\,f=-8\pi G_N T_{UU},\,\, T_{UU}=p^V\left(\frac{F^2}{\kappa^2UV}\right)_{U_0}\delta(\theta-\pi/2)
\label{backreaction_diff_axi_symmetric}
\ee
Here $\mathcal{D}_{\theta}$ is a first order differential operator of rank 2. 
We normalize $f$ by choosing $f(\pi/2)=1$, thus $\alpha$ captures the strength of the back reaction while $f(\theta)$ captures the profile of the perturbation around the sphere. Here $p^V$ denotes the strength of such a shockwave. We will determine the value of $\alpha$ by demanding smoothness along the transverse volume element as the perturbation tends to the past horizon (past from the right exterior $pov$).
\\\\
The problem we want to analyse is the one studied first in \cite{Shenker:2013pqa} wherein one sends an infinitesimally small ($\mathcal{O}(G_N)$) perturbation into the black hole from the left exterior at a time $t_0$ which is in the past of $t=0$ slice of the right exterior. This perturbation slowly grows in strength as it falls into the black hole and at late time develops enough energy (as measured locally) that the back reaction of the metric is of $\mathcal{O}(1)$. It is useful to work with $\{\tau,z\}$ coordinates, as $\tau=t-\Ll\phi$ for very large times $t_0\gg\beta$ implies $\tau_0\gg\beta$. The Kerr geometry is also periodic in terms of the co-moving coordinate $z=\phi-\mu \,t$. We therefore parametrize the axi-symmetric shockwave by $U=U_0$
\be
U_0=e^{-\kappa \tau_0}
\ee 
where $\tau_0\gg\beta$ is the time on the left boundary. 
As $\tau_0\rightarrow\infty$ $U_0\rightarrow 0$ therefore the above Dray-'t Hooft solution is recovered in this limit. It is important to note that the Dray-'tHooft solution can only describe the late time behaviour of the back-reacted metric. 
Imposing smoothness of the  volume element $H=\sqrt{{\rm det}[g_{\{\theta,z\}}]}$ transverse to the shockwave at $U=U_0$ implies
\be
H \big\vert_{U_0^+} = H\big\vert_{U_0^-}
\ee
Where we use $\widetilde{V}$ coordinate for $U>U_0$ and $V$ for $U<U_0$. At late times $i.e.$ as $\tau_0\rightarrow \infty$ we expect $\widetilde{V}$ to be shifted as compared to $V$ by a step function. We determine the magnitude of this shift by demanding the above smoothness condition holds as we approach the Dray-'t Hooft solution. 
\be
H=H_0+H_1(r-\rpl)=H_0-H_1\left(\mathbb{A} \frac{U_0 V}{F'(\rpl)}\right)
\ee
Denoting parameters after the shockwave with a tilde we have
\bea
&&H_0-H_1\left(\mathbb{A} \frac{U_0 V}{F'(\rpl)}\right)=\widetilde{H}_0-\widetilde{H}_1\left(\widetilde{\mathbb{A}} \frac{U_0 \widetilde{V}}{\widetilde{F}'(\rpl)}\right)\cr&&\cr
\implies&&\widetilde{V}=\frac{\widetilde{F}'_+\mathbb{A}H_1}{F'_+\widetilde{\mathbb{A}}\widetilde{H}_1}V+\frac{(\widetilde{H}_0-H_0)\widetilde{F}'_+}{\mathbb{A}\widetilde{H}_1U_0},\hspace{0.3cm}{\rm where}\,\,F'_+\equiv F'(\rpl)
\label{Shift_V}
\eea
Note that $H_0$ is simply the horizon area divided $4\pi^2$. Therefore we expect $\delta H_0$ to vary according to the first law as 
\be
\delta H_0 \sim \beta(\delta \mathcal{M}-\mu\, \delta J)=\delta \mathcal{S}
\ee
where $\mathcal{M}$ and $J$ are the ADM mass and angular momentum of the black hole computed appropriately \cite{Papadimitriou:2005ii}.
\be
\mathcal{M}=\frac{M}{\Xi^2},\hspace{0.3cm}J=\frac{Ma}{\Xi^2}
\label{ADM_M_J}
\ee
We work in the limit where the black hole has large entropy, thus evaluating $\frac{F'_+}{H_1}$ in the limit $\mathcal{S}\rightarrow \infty$ we find
\be
\frac{F'_+}{H_1}=\frac{\mathbb{B}}{\mathcal{S}}
\ee
where $\mathbb{B}$ depends only on $\rmi/\rpl$ and $\theta$.
We further have the  limit $U_0\rightarrow 0$ which is simply the limit that the perturbation was sent in the far past as $\tau_0\rightarrow \infty$ on the boundary of the left exterior implies far past $w.r.t.$ the CFT on the right boundary. We consider the following scaling in this limit with $\mathcal{S}\rightarrow \infty$
\be
U_0\rightarrow 0,\hspace{0.3cm}\frac{(\delta\mathcal{M}-\mu\, \delta J)}{T_H \mathcal{S}}\rightarrow 0,\hspace{0.3cm}\frac{(\delta\mathcal{M}-\mu\, \delta J)}{U_0\,T_H \,\mathcal{S}}\rightarrow{finite}.
\label{limit}
\ee
In this limit the above shift in $\widetilde{V}$ \eqref{V_shift} has a finite contribution. The perturbation we consider further has a specified angular momentum per unit energy $\Ll$, this implies that if the change in black hole's ADM mass is $\delta\mathcal{M}\sim E_0$ - $E_0$ being the energy of the shockwave measures at the $AdS$ boundary; then $\delta J=\delta \mathcal{M}\mu\Ll\sim E_0\mu\Ll$. Therefore we have
\be
\delta\mathcal{M}-\mu\delta J=\delta\mathcal{M}(1-\mu\,\Ll)
\ee
Absorbing the non-extensive parameters $(\mathbb{B},\mathbb{A},2\pi)$ in the variation of ADM mass and angular momentum we can write
\be
\widetilde{V}=V-\frac{\Theta(U-U_0)}{U_0}\frac{\delta\mathcal{M}(1-\mu\,\Ll)}{T_H \mathcal{S}}
\label{V_shift_3}
\ee
where we have introduced the step function to indicate the shift in $V$ across the shock-wave.
\\\\
Working with the above limits imply  our analysis holds for large black holes with large entropy as compared to the perturbation.
It is worth mentioning that the shift $\delta\mathcal{M}$ has complicated $\theta$ dependence and  is further supposed to solve a differential equation implied by \eqref{backreaction_diff_axi_symmetric}. This should not be surprising as the geometry after the shockwave is no longer a stationary solution to Einstein's equations hence the $\delta\mathcal{M}$ can and must have spatial dependence\footnote{The time dependence comes $via$ the dependence on $V$ and would play an important role if one relaxes the limit $U_0\rightarrow 0$. }. Comparing this with  \eqref{V_shift_0} we see that 
\be
\alpha f(\theta)=\frac{\beta\,\delta\mathcal{M}(1-\mu\, \Ll)}{U_0\,\mathcal{S}}=\frac{\beta (1-\mu\,\Ll)E_0}{U_0\,\mathcal{S}}f(\theta)
\label{alpha_def}
\ee
where the transverse $\theta$ dependence on the right is captured by $\delta\mathcal{M}$. We have taken $\delta\mathcal{M}$ to be proportional to the total energy $E_0$ of the perturbation measured at the boundary at  left exterior time $\tau_0$. Therefore we have
\be
\alpha=\frac{1}{U_0}\frac{\beta (1-\mu\,\Ll)E_0}{\mathcal{S}},\hspace{0.3cm} {\rm and }\hspace{0.3cm} V\rightarrow \widetilde{V}=V-\frac{1}{U_0}\frac{\beta(1-\mu\,\Ll) E_0}{\mathcal{S}}f(\theta) 
\label{V_shift}
\ee 
The inverse $U_0^{-1}=e^{\kappa \tau_0}$ dependence is indicative of the perturbation growing at an exponential rate dictated by $\kappa\geq T_H$ for an arbitrary value of angular momentum $\mathcal{L}\geq 0$. 
We can in principle solve for $f(\theta)$ to find the how the backreaction spreads away from the equator; a similar analysis can also be done for a point null particle perturbation at the equator. This must indeed reveal interesting dynamics and butterfly velocity associated with such perturbations to the rotating geometry; see \cite{Blake:2021hjj} for a analysis of the butterfly velocity in a slowly rotating Kerr $AdS_4$. However  for the case at hand we need not be concerning ourselves with dynamics away from the equator and choose $f(\theta=\pi/2)=1$.   
\\\\
In the above analysis we have assumed that the shockwave emanates at the boundary of $AdS_4$ at a near horizon time $\tau_0$, the $\{\tau,z\}$ coordinates are related to the Boyer-Lindquist coordinates $\{t,\phi\}$ by \eqref{tau_z_to_t_phi}. 
The choice of the $\{\tau,z\}$ coordinates is forced upon us from arguments presented subsection 3.1.  It is important to note that demanding the smoothness of the Dary-'tHooft solution at the outer horizon we are forced to work with the periodic coordinate $z$. 
Had we been analysing the backreaction due to a single in-falling particle localized in $z$ the resultant response function $f(\theta,z)$ would have to periodic in $z$ and not in $\phi$. 
One way to understand this is that $z$ is the co-moving coordinate outside the horizon and in the ergo-region where no stationary observer can exist who is time like. 
Since the response function has to periodic in $z$ it cannot grow exponentially in $z$ (with a real coefficient). 
Therefore $\alpha f(\theta,z)$ occurring instead in \eqref{alpha_def} would still have an exponential behaviour only in $\tau_0$.
\\\\
For the case at hand of a shockwave existing at every point in the $z$ coordinate and starting out from the $AdS_4$ boundary at $\tau_0$ we would have a spread in the time $t^{(s)}=\left(1-\frac{a\Ll}{\ell^2}\right)t$ of the static boundary coordinate given by \eqref{tau_z_to_t_phi} 
\be
\tau_0=t_0^{(s)}-\Ll\,\phi_0, \hspace{0.3cm}\forall \,\,\phi_0\in [0,2\pi]
\ee
and the spread in $[0,2\pi]$ in $\phi$ ensures an equivalent spread in $z$ by the time the shockwave reaches the outer horizon\footnote{This is because we demanded $\int_0^{2\pi}\sqrt{-g}d\phi=\int_0^{2\pi}\sqrt{-g}dz$ in the near horizon region in defining $z$. As the backreaction is maximum  closer to $\rpl$ it is this periodicity which is the most relevant for the Dray-'tHooft solution.  }. 
Thus fixing a fixed time $\tau_0$ at the boundary for the start of perturbation implies sending the first pulse at $\phi=0$ at $t_0^{(s)}=\tau_0$. We will henceforth  mention only the behaviour with regards to $\tau_0$ as this signals the start of the perturbation and linear shifts of the form indicated above would not affect the qualitative statements made towards the end.
\section{Extremal Surfaces}
\begin{figure}
	\centering
	\includegraphics[scale=0.2]
	{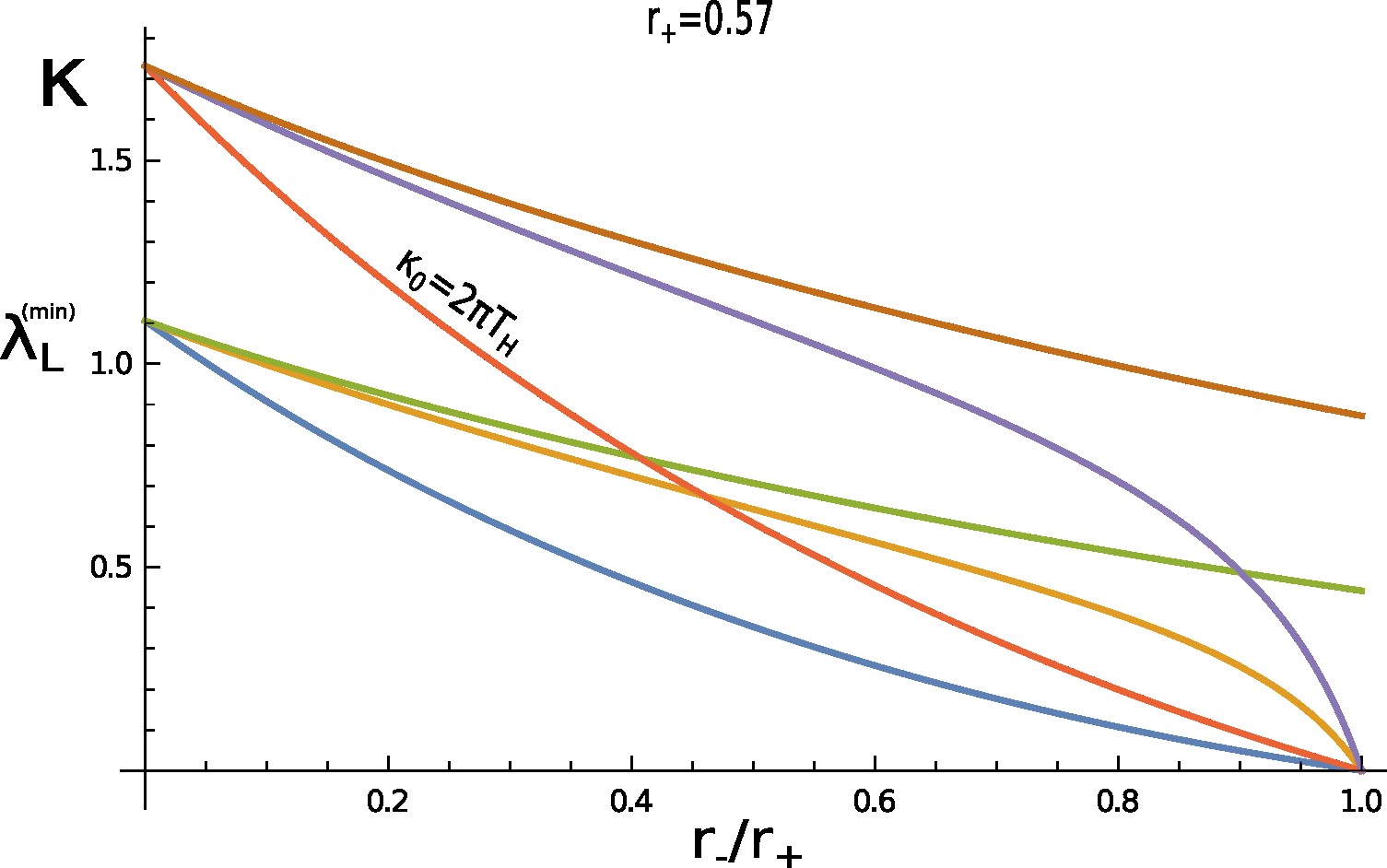}
	\caption{Plots for various values of $\Ll=s \left(\frac{\rmi}{\rpl}\right)\mu^{-1}$ with $\rpl=0.57$ and $\ell=1$. The blue, orange and green curves are for$\lambda^{(\rm min)}_L$ corresponding to $s=\{0,\frac{9}{10},1\}$ respectively while the red, violet and brown curves correspond to the value of $\kappa$ for similar respective values of $s$. The red curve $\kappa_0$ $i.e.$ the temperature of the geometry is greater than only the $\lambda^{(\rm min)}_L$ for $\mathcal{L}=0 \,(s=0)$; the values of $\lambda^{\rm (min)}_L$ for $\mathcal{L}>0$ can be greater than this curve.  }
\label{F_cH_c_kappa}
\end{figure}

We next compute the disruption of Mutual Information between the TFD states due to such a shockwave along the equator. We consider a hemispherical subsystem  with the equator at $\theta=\pi/2$ as it's boundary, and consider identical subsystems $A$ and $B$ at both the $left$ and $right$ Kruskal boundaries. The Mutual Information is then given by
\be
I[A:B]=S_A+S_B-S_{A \cup B }
\ee
with the appropriate HRT surface used to compute the $rhs$ above \cite{Hubeny:2007xt}. The shockwave geometry is locally unchanged from the Kerr metric in each of the exteriors in the Kruskal diagram except for the shift in the ingoing $V$ coordinate \eqref{V_shift} along the past horizon denoting the change due to the perturbation at very late times. Observe that only the HRT surface  corresponding  to  $S_{A\cup B}$  traversing the geometry from $right$ to $left$ is sensitive to this shift and  hence detects the effect of this shockwave. It can be explicitly shown that the HRT surface dual to $S_{A\cup B}$ lies in the equatorial plane as it is a co-dimension two extremal surface in both the exteriors. 
\\\\
The extremal surface inherits the axi-symmetry of the Kerr geometry along the $z$ coordinate. We 
extremize the surface $w.r.t.$ the co-ordinates $\{r,\tau,z\}$ and the metric given in \eqref{Kerr_LC_equator}. 
As the extremal surface is symmetric in the $z$ coordinate we only need to extremize it along the $\{r,\tau\}$ direction. We would only be interested in the late time behaviour of the dependence of this extremal surface on the shift in the $V$ coordinate. The procedure for obtaining  this change in the extremal surface would be identical to the one followed in \cite{Leichenauer:2014nxa}.  
\\\\
The subsystem we consider is the union of upper hemispheres at the 2 boundary $CFT_3$s and the corresponding extremal surface homologous to it is the $\theta=\frac{\pi}{2}$ surface for which it can be shown that the trace of extrinsic curvature vanishes\footnote{Here we consider the trace of the extrinsic curvature constructed out of the unit normal in the $\theta$ direction. The HRT surface further needs to be extremised in the time direction $\tau$.}. The induced metric on $\theta=\frac{\pi}{2}$ surface given by \eqref{Kerr_LC_equator} can be expressed in  $\{r,\tau\}$ coordinates  as follows 
\be
ds^2_{\theta=\tfrac{\pi}{2}}=F(-d\tau^2+\frac{dr^2}{f^2})+ h\,(dz+h_\tau d\tau)^2 
\label{Kerr_Kruskal_equatorrtau}
\ee
 This implies that the area of the extremal surface ${\cal A}_{\theta=\pi/2}$ is given by extremizing the following functional
\begin{align}\label{area}
{\cal A}_{\theta=\tfrac{\pi}{2}}=S_{z} \int d \tau \sqrt{h} \sqrt{-F+F f^{-2} \dot{r}^{2}}
\end{align}
where $S_{z}=2\pi$ corresponds to the transverse volume obtained by integrating along $z$ direction. Note that the area functional is independent of $\tau$ coordinate which leads to a  conserved quantity given as
\begin{align}\label{com}
\gamma=\frac{-F \sqrt{h}}{\sqrt{-F+F f^{-2} \dot{r}^{2}}}=\sqrt{-F_{0}h_0}. 
\end{align}
\begin{figure}
	\centering
	\includegraphics[scale=0.15]{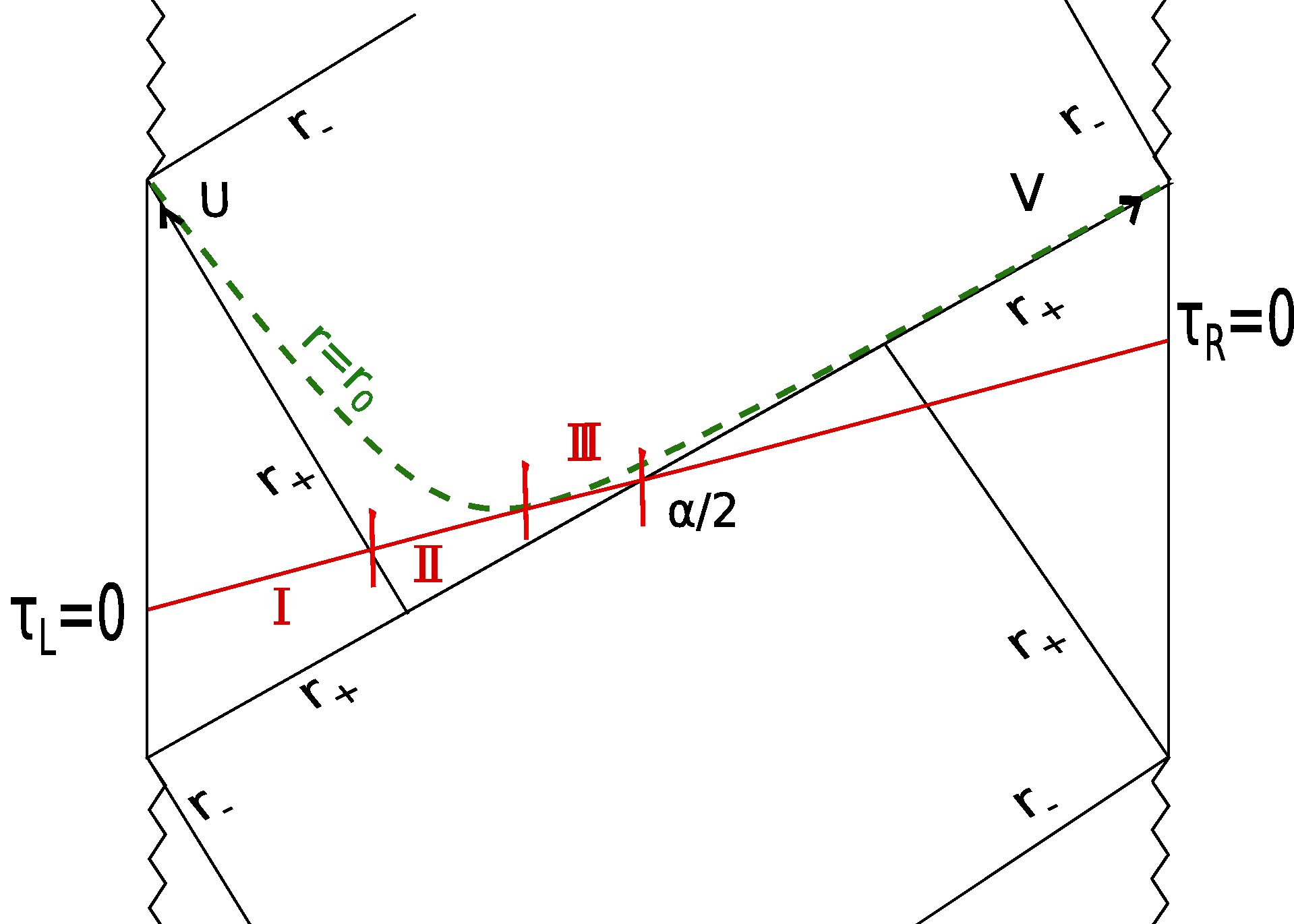}
	\caption{Figure depicting the extremal surface (Red) at $t=0$ in extended Kerr spacetime which is analysed by dividing into three segments denoted as I, II and III. The shift $V$ coordinate at $U=0$ is $\alpha$. Note, the lengths II and III are the same and so are the lengths along the red curve on the either sides of $U=0\,\,\&\,\,V=\alpha/2$. }
\label{KerrExt}
\end{figure}
The above expression can also be found by solving for $\dot{r}$ from its $eom$. $\gamma$ is thus a constant  that describes the extremal surface.  Here 
 $H_0$ and $F_0$ are simply the values of function $H,F$ at a point where $\dot{r}=0$. 
Following \cite{Leichenauer:2014nxa} we would try to learn about the nature of the extremal surface from the above expression from it's dependence on $\dot{r}$.\\\\ 
Observe that $F_0$ is considered to be negative as $r_0$ lies behind the horizon. When $r_0$ approaches horizon $i.e.$ $r_0 \to r_+$, $\gamma \to 0$ (as $F\sim\mathcal{O}(r-\rpl)$) which is also the limit $\alpha\to 0$ $i.e.$ when the shock wave is absent.
Note that the  \eqref{com} can be inverted to find $\tau$ as a function of $r$ as follows
\begin{align}
\tau(r)=\int dr \frac{1}{f\sqrt{1+\gamma^{-2}Fh }} 
\end{align}
The tortoise coordinate $r_*$ relevant for our analysis can also be obtained as a function of $r$ through the following integral
\be
 	\quad r_*(r)=-\int_\infty^r  \frac{dr'}{f(r')},
\ee 
however the exact form of this integral would not be needed. It suffices to state that the $AdS$ boundary occurs at $r_*=0$ and the outer horizon at $r_*=-\infty$ for non-extremal black hole parameters.
\\\\ 
Following  \cite{Leichenauer:2014nxa} we compute the area of the extremal surface by dividing its left half  three parts as depicted in Figure-\ref{KerrExt} and finally we obtain the full area by symmetry  which is twice that of the left half. Note, when $\alpha=0$ the above figure would not have a shift along the $V$ coordinate $i.e.$ the $V=0$ surface would be a continuous straight line. Further the extremal surface in red would be a horizontal straight line beginning at $t_R=0$ at the right boundary, passing through the bifurcate horizon $(U=0=V)$ and ending at $t_L=0$ on the left boundary. Let us first consider the segment I- starting from the boundary $(U,V)=(1,-1)$ to $(U,V)=(U_1,0)$  the point where extremal surface intersects $V=0$. For this surface we have
\begin{align}
U_1^2&=\exp{\bigg[2 \kappa (dr_*-d\tau)\bigg]}\nonumber\\
&=\exp\bigg[-2\kappa \int_{r_+}^{\infty} \frac{dr}{f}\bigg(1- \frac{1}{\sqrt{1+\gamma^{-2}Fh }}\bigg)\bigg]
\end{align}
The second segment II begins at $(U,V)=(U_1,0)$ and  ends at  $(U_2,V_2)$ which lies on the surface $r=r_0$ which is the turning point for the extremal surface. Here we have
\begin{align}
	\frac{U_2^2}{U_1^2}=\exp\bigg[-2\kappa \int_{r_0}^{r_+} \frac{dr}{f}\bigg(1- \frac{1}{\sqrt{1+\gamma^{-2}Fh }}\bigg)\bigg]
\end{align} 
This in turn leads to
\begin{align}
	U_2^2=\exp\bigg[-2\kappa \int_{r_0}^{\infty} \frac{dr}{f}\bigg(1- \frac{1}{\sqrt{1+\gamma^{-2}Fh }}\bigg)\bigg]
\end{align}
In order to find $V_2$ as  in \cite{Leichenauer:2014nxa} we consider a reference surface described by $r=\bar{r}$ at which $r_*=0$ and hence we get
\begin{align}
	V_2=\frac{1}{U_2}\exp\bigg[-2\kappa\int^{r_0}_{\bar{r}}\frac{dr}{-f}\bigg]
\end{align}
Similarly considering the segment III we find $V_2$ in terms of the intersect at $U=0$ and $V=\alpha/2$ as 
\begin{align}
	\frac{\alpha^2}{4 V_2^2}=\exp\bigg[2\kappa\int_{r_0}^{r_+}\frac{dr}{f}\bigg(1+\frac{1}{\sqrt{1+\gamma^{-2}Fh }}\bigg)\bigg]
\end{align}
Hence $\alpha$ is described by a combination of three integrals as follows
\begin{align}
	\alpha= 2 \exp\bigg(Q_1+Q_2+Q_3\bigg)
\end{align} 
where the integrals $Q_1,Q_2$ and $Q_3$ take the form
\begin{align}
	Q_1&=-2\kappa\int^{r_0}_{\bar{r}}\frac{dr}{-f}\\
	Q_2&=2\kappa \int_{r_0}^{\infty} \frac{dr}{f}\bigg(1- \frac{1}{\sqrt{1+\gamma^{-2}Fh }}\bigg)\\
	Q_3&=2\kappa\int_{r_0}^{r_+}\frac{dr}{f}\bigg(1+\frac{1}{\sqrt{1+\gamma^{-2}Fh }}\bigg)\label{Q3}
\end{align}
Observe that integrals $Q_1$ and $Q_2$ diverge as $r_0\to r_+$ which is also the limit in which $\alpha\to0$. Notice that $Q_3$ diverges as $r_0\to r_{c}$ where $r_c$ corresponds to the point at which the following expression holds
\begin{align}
h(r_c)F'(r_c)+h'(r_c)F(r_c)=0
\label{r_c_equation}
\end{align}
The above equation is cubic in $r$ and $r_c$ can be obtained analytically.
However, notice that $Q_1$ and $Q_2$ are finite as $r_0\to r_{c}$. Hence as $r_0\to r_c$, $Q_3$ alone diverges and corresponds to $\alpha\to \infty$. Let us now try to compute the divergent part of the area in \eqref{area} which can be re-expressed using \eqref{com} as follows
\begin{align}
{\cal A}_{\theta=\tfrac{\pi}{2}}=2\pi \int \frac{d r}{f} \frac{ F h/\gamma}{\sqrt{1+\gamma^{-2}Fh }}
\end{align}
Note that at late times when $\alpha$ becomes large the dominant contribution to the above extremal surface area comes from its segment near $r=r_c$ and in this regime the area above can be approximated to be proportional to the $Q_3$ integral given in \eqref{Q3}. Note that the total area contributing to the extremal surface corresponding to  $S_{A\cup B}$ is given by four times the segment evaluated above which gives the dominant contribution for large $\alpha$. This is expressed as
\begin{align}
{\cal A}_{A\cup B}&\approx\frac{4\pi}{\kappa}\sqrt{-F_ch_c}\,Q_3
=\frac{4\pi}{\kappa}\sqrt{-F_ch_c}\,\log\alpha
\end{align}
Upon substituting the expression for $\alpha$ we derived in \eqref{V_shift}, the above result for the extremal area reduces as follows
\begin{align}\label{AreaAB}
{\cal A}_{A\cup B}&\approx 4\pi\tau_0\sqrt{-F_ch_c} \,+\frac{4\pi}{\kappa}\sqrt{-F_ch_c} \,\log\bigg[\frac{\beta E_0\,(1-\mu\Ll)}{\mathcal{S}}\bigg]
\end{align}
The above equation clearly indicates that the rate of growth of area of the extremal surface corresponding to $S_{A\cup B}$ $w.r.t.$ the time-stamp $\tau_0$ of the perturbation is controlled by the parameter $\sqrt{-F_ch_c}$.
Also note that the above expression is only valid for large $\alpha$ and hence the unperturbed value of the area $\mathcal{A}_{\theta=\frac{\pi}{2}}$ cannot be inferred from this expression.  
\\\\
Let us pause here to compare the above result for generic values of $\mathcal{L}$. Note that for $\Ll< \mu^{-1}$ we have $F\sim \mathcal{O}(r-\rpl)$. However only for $0\leq\Ll<\mu^{-1}$ we have $F'(\rpl)\rightarrow 0$ as $\rmi\rightarrow\rpl$ $i.e.$ as the configuration approaches extremality. This implies that for such shockwaves $r_c\rightarrow\rpl$ as the condition \eqref{r_c_equation}  is satisfied as we take the configuration arbitrarily close to extremality. Let us compare this with the case where $\Ll=\frac{\rmi}{\rpl}\mu^{-1}$ where $\Ll\rightarrow\mu^{-1}$ as we approach an extremal configuration. For such a shockwave the condition \eqref{r_c_equation} is not similarly satisfied  at $r=r_c=\rpl$ and consequently $F_c$ does not approach zero as $\rmi\rightarrow\rpl$. 
In order to compare quantitatively, it is useful to divide the above expression for $\mathcal{A}_{A\cup B}$  by a quantity which behaves similarly in terms of  black hole parameters for $0\leq \Ll<\mu_{-1}$, we choose this to be the unperturbed value for a similar area $\mathcal{A}^{(0)}_{A\cup B}$. However since the exact analytic expression for the unperturbed $\mathcal{A}^{(0)}_{A\cup B}$ is not easily tractable, we utilize the relation that  for large enough subsystems, entanglement entropy scales like the entropy of the system. Hence, we take it to be proportional to the black hole's entropy 
\be
\mathcal{A}^{(0)}_{A\cup B}=\rho\, \mathcal{A}_{H},\hspace{0.4cm}{\rm where}\,\,\rho\in [0,1].
\label{A_unperturbed}
\ee
Therefore we can write the term proportional to $\tau_0$ in \eqref{AreaAB} as
\be
\mathcal{A}_{A\cup B}=\mathcal{A}^{(0)}_{A\cup B}\,\lambda_L\,\tau_0\implies\lambda_L=\frac{4\pi\sqrt{-h_cF_c}}{\rho \,\mathcal{A}_{H}}
\label{lambda_inst}
\ee
\begin{figure}
	\centering
	\includegraphics[scale=0.2]
	{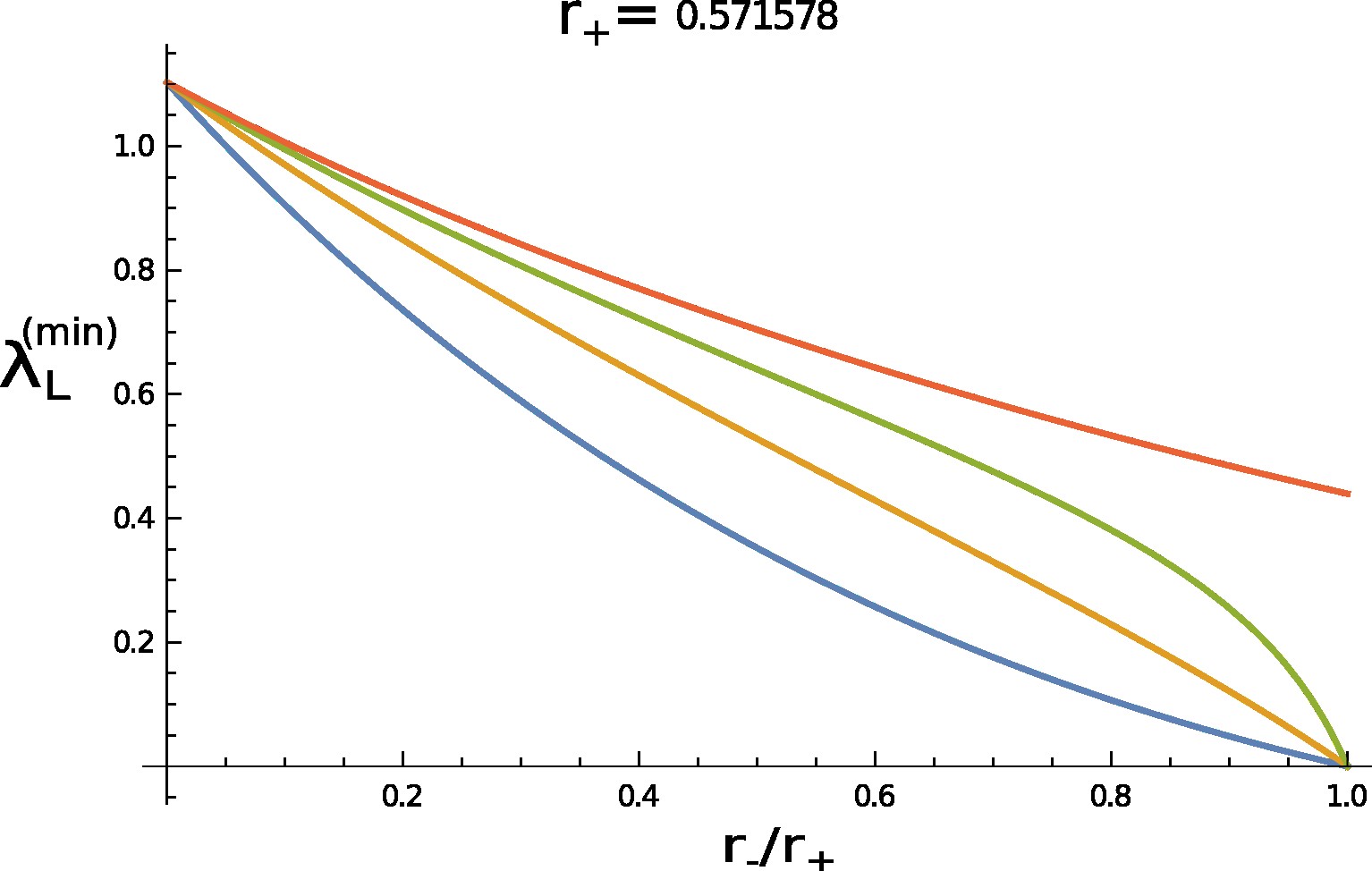}
	\caption{Plots for various values of $\Ll=s \left(\frac{\rmi}{\rpl}\right)\mu^{-1}$ with $\rpl=0.57$ and $\ell=1$. The blue, orange and green curves correspond to $s=\{0,\frac{2}{3},\frac{9}{10}\}$ respectively. These curves approach zero as $\rmi\rightarrow\rpl$. The red curve corresponds to $s=1$ and approaches a constant since $\Ll\rightarrow\mu^{-1}$ as $\rmi\rightarrow\rpl$. }
\label{F_cH_c}
\end{figure}
\begin{flushleft}
where $\lambda_L$ can be regarded as the instantaneous Lyapunov exponent at late times $i.e.$ $\gg \beta^{-1}$.
For $\rho=1$ we get the minimum possible instantaneous $\lambda_L$
\end{flushleft}
\be
\lambda^{\rm (min)}_L=\frac{4\pi\sqrt{-h_cF_c}}{ \,\mathcal{A}_{H}}.
\label{lambda_min}
\ee
We plot this for different values of $\Ll$ in Fig.\eqref{F_cH_c}. Here we do not plot the values exactly at the extremal point. It is clear from the above plot that if one were to choose an angular momenta $\Ll$ for the shockwave such that it approaches the value $\Ll\rightarrow\mu^{-1}$ as the configuration approaches extremality, then the rate of scrambling does not approach zero. Note, away from extremality one can not choose $\Ll=\mu^{-1}$ as this value is not allowed by turning point analysis $i.e.$ the shockwave does not reach the horizon. 
We can also compare the value of $\lambda^{(\rm min)}_L$ to the temperature $\kappa_0=2\pi T_H$ and  $\kappa$ which dictates the blueshift suffered by the in-falling null rotating particle generating the shockwave. We plot these in Fig \eqref{F_cH_c_kappa} against the ratio $\rmi/\rpl$. It is apparent that the instantaneous $\lambda_L^{(\rm min)}$ seems to be bounded effectively by $\kappa$. One can observe that temperature $\kappa_0=2\pi T_H$ indicated by the red curve can be less than $\lambda^{\rm (min)}_L$ when the geometry is sufficiently non-extremal. Thus clearly suggesting that the instantaneous rate of scrambling can be greater than the temperature of the Kerr geometry but bounded by the exponent $\kappa$ determining the blueshift for a rotating null in-falling  shockwave.
\\\\
Having obtained the rate of growth of the HRT surface let us now try to estimate the scrambling time through mutual information given by
\begin{align}
	I(A:B)=S_A+S_B-S_{A\cup B}.
\end{align}
Note that for large enough $\tau_0$ or $\alpha$,  the extremal surfaces corresponding to the individual subsystems $A$ and $B$ remain unaffected by the change in the geometry. 
\begin{align}
	I(A:B)=S_A+S_B-\frac{{\cal A}_{A\cup B}}{4G_N}
\end{align}
where $S_A,S_B$ are the unperturbed entanglement entropies of the subsystems $A$ and $B$.  
Hence, using \eqref{AreaAB} we have
\begin{align}
	I(A:B)\approx S_A+S_B-\frac{\pi\tau_0}{G_N}\sqrt{-F_ch_c} \,-\frac{\pi}{\kappa G_N}\sqrt{-F_ch_c} \,\log\bigg[\frac{\beta (1-\mu\,\Ll) E_0}{\mathcal{S}}\bigg].
\end{align}
We will take the energy $E_0$ of the perturbation to be that of the order of the energy of a few Hawking quanta, thus $\beta E_0\sim 1$. Further, as the turning point analysis had revealed \eqref{turning_point} for non-extremal configurations $\Ll_{max}<\mu^{-1}$, therefore $0<(1-\mu\,\Ll)<1$ and  does not scale extensively as compared to the the black hole's entropy\footnote{This quantity can be large but would only depend on how one chooses the approach of $\Ll$ to $\mu^{-1}$ or a fraction of $\mu^{-1}$, as the geometry tends to extremality. It would not therefore depend on the size of the black hole.}.
This in turn leads to the following expression for the scrambling time
\begin{align}\label{sc1}
 \kappa\tau_*\approx \log\mathcal{S}+ \frac{\kappa G_N(S_A+S_B)}{\pi\sqrt{-F_ch_c}} + \log\frac{1}{1-\mu\,\Ll}
\end{align}
\begin{figure}
	\centering
	\includegraphics[scale=0.2]
	{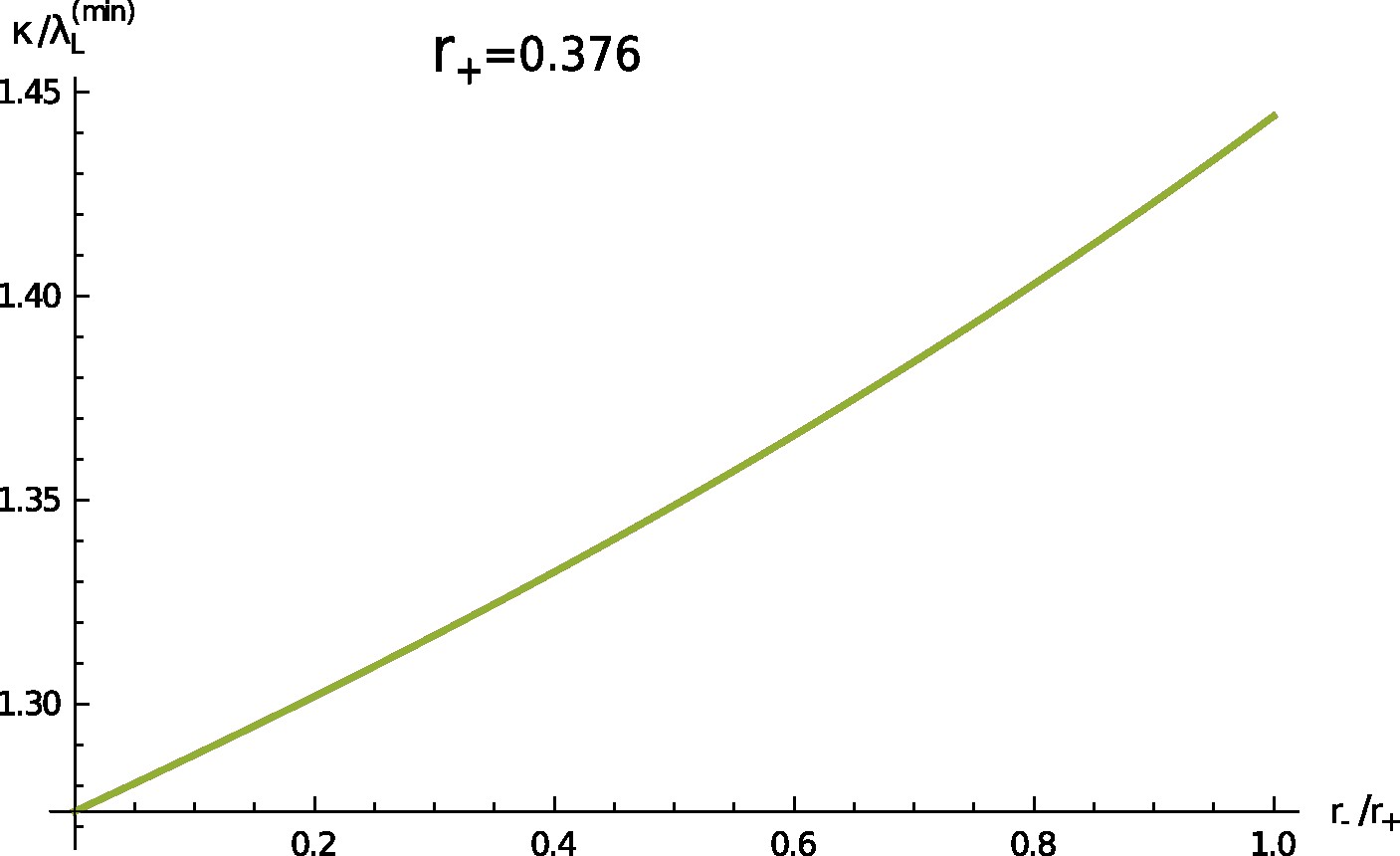}
	\caption{Plot of $\kappa/\lambda^{(min)_L}$ against $\rmi/\rpl$ with $\ell=1$. This plot is independent of the shockwaves specific angular momentum $\Ll$. }
\label{kappa_by_lambda_min}
\end{figure}
which can be re-expressed as 
\begin{align}
\kappa\tau_*\approx\log\mathcal{S}\,+\, \frac{\kappa ({\cal A}_A+{\cal A}_B)}{4\pi\sqrt{-F_ch_c}} \,+\, \log\frac{1}{1-\mu\,\Ll}
\end{align}
where ${\cal A}_A+{\cal A}_B$ are the unperturbed areas of the  extremal surfaces homologous to the subsystems $A$ and $B$ respectively computed at late times. Notice that as all the terms in the above expression are dimensionless, the second term is non-extensive (${\cal A}_A$ and $\sqrt{-F_ch_c}/\kappa$ both have dimensions of area.) $i.e.$ it does not scale with the entropy of the system
\be
\frac{\kappa(\mathcal{A}_A+\mathcal{A}_B)}{4\pi\sqrt{-F_ch_c}}=\frac{\kappa}{\lambda_L^{(min)}}\frac{(\mathcal{A}_A+\mathcal{A}_B)}{\mathcal{A}_{H}}\approx\kappa\frac{S_A+S_B}{S_{A\cup B}}
\ee
where we used \eqref{lambda_min} and $\kappa/\lambda^{(min)}_L$ although complicated is found to be independent of $\Ll$ $c.f.$ Fig-\ref{kappa_by_lambda_min}. It is also worth noting that this term survives the limit $\rmi\rightarrow 0$ in Fig-\ref{kappa_by_lambda_min}, therefore it exists even for the case of static black holes in $AdS_4$.
\\\\
The third term in \eqref{sc1} does depend on $\mathcal{L}$ and is positive. This term only depends on the chemical potential and the angular momentum (per unit energy of the shockwave) $\Ll$ and has a tendency to increase the scrambling time. However as this term -like the second, does not scale with an increasing function of the black hole entropy $\mathcal{S}$ the effect of such a decrease is parametrically small for large black holes. This term has recently made an appearance in the charged shockwave analysis in RN-$AdS_4$ \cite{Horowitz:2022ptw} where the effect of delay in the scrambling time due to a charged massless shell of in-falling matter was obtained to be
\be
t_*^{(q)}-t_*=\frac{\beta}{2\pi}\log \left(1-\mu \frac{q}{E_0}\right)^{-1}
\label{RN_charged_sc1}
\ee 
where the $lhs$ is the difference in the scrambling times of a charged shockwave to an uncharged one. Here $\mu$ and $\beta^{-1}$ is the chemical potential and temperature of the RN-$AdS_4$ respectively and $q/E_0$ -like $\Ll$, is the charge per unit energy of the shockwave. As was shown in section 2 (appendix \ref{RN}) the Lyapunov index for RN-$AdS_4$ is not expected to change if the shockwave has an angular momentum and is given by the temperature of the static geometry.
\\\\
For large black holes where $\log\mathcal{S}$ is much larger than the rest of the terms in \eqref{sc1}, the scrambling time is given by
\begin{align}
\tau_*\approx \frac{1}{\kappa}\log\mathcal{S}
\label{scrmabling_time}
\end{align}
Although the rate of growth of the extremal surface for the $A\cup B$ subsystem is controlled by the parameter $\sqrt{-F_ch_c}$ as described earlier, we observe from the above equation that the scrambling time is controlled by $\kappa$. Thus $\kappa$ can be regarded as the rate of growth of the disruption of mutual information at very large time scales $i.e.$ when $I(A:B)\rightarrow 0$. This time scale is larger than the late times at which $\lambda^{(min)}_L$ was computed for.
\\\\
We must also note that  the above analysis is applicable for generic non-extremal black holes and the behaviour of scrambling close to zero temperature for large black holes is determined by their near extremal limit. Here the wormhole connecting the two asymptotic boundaries becomes infinitely long, in contrast with the strict extremal case wherein we have a disconnected boundary.  
\section{Conclusions \& Discussions}
We study the butterfly effect for rotating geometries in $AdS_4$ by computing systematically the rate of disruption of mutual information at very late times due to in-falling rotating shockwaves with angular momenta per unit energy $\Ll$. We find like the analysis in \cite{Shenker:2013pqa} that this rate is controlled by the blueshift suffered by the in-falling shockwave which is given by $\kappa$ \eqref{kappa_1} which for angular momentum $\Ll>0$ can be greater than the temperature of the black hole $\kappa_0=2\pi T_H$. We also find that the rate of growth of the wormhole $i.e.$ the HRT surface connecting the two boundaries, as a response to the shockwave can grow at rate $\lambda_L^{\rm (min)}$ which can survive the extremal limit for a particular values of angular momentum $\Ll$ Fig\eqref{F_cH_c}. This is in contrast with similar analysis in the static case or for the case with $\Ll=0$ where this rate drops to zero and is always bounded by the temperature of the black hole. We also find that $\lambda^{\rm (min)}_L$ can easily be greater than the temperature of the black hole for $\mathcal{L}>0$ for sufficiently non-extremal geometries. This rate is at best bound by the blueshift $\kappa$ even when it survives the extremal limit. The scrambling time also behaves as $t_*\sim\kappa^{-1}\log \mathcal{S}$ in such cases. 
We also find terms that increase the scrambling time but these terms do not scale like the black hole entropy for large black holes. We find one such term which increases the scrambling time to be $\log(1-\mu\,\Ll)^{-1}$ which expressly depends on the angular momentum $\Ll$ of the shockwave. Such a delay has recently been observed for the case of charged shockwaves in RN-$AdS_4$ \cite{Horowitz:2022ptw}.
Interestingly for near extremal black holes as $\kappa$ survives this limit for values of $\Ll$ which tend to $\Ll\rightarrow \mu^{-1}$,  the scrambling time is still proportional to the extremal degrees of freedom of the Kerr geometry $i.e.$ $t_*\sim\kappa_{ext}^{-1}\log \mathcal{S}_{ext}$ \eqref{extremal_blueshift}. This particular feature is very interesting given that string theory is able to rightly predict the microscopic degeneracy of extremal \cite{Strominger:1996sh} and near extremal \cite{Maldacena:1996ds} super-symmetric black holes.
\\\\
There are some interesting features of the above result when contrasted with a similar study of scrmabling of mutual information due to rotating shockwaves in BTZ \cite{Malvimat:2021itk}. In it the turning point analysis allowed for $\Ll=1=\mu^{-1}_{ext}$, which is a peculiarity of the BTZ as for higher dimensional black holes $\Ll<\mu^{-1}$. The scrambling time and the Lyapunov index in BTZ can be computed analytically for such a case  \cite{Malvimat:2021itk} and was found to be controlled by $\lambda_L=\kappa/2=\frac{\pi T_H}{1-\mu}=\rpl$. In contrast here we find that the scrambling time is controlled by $\kappa$ \eqref{scrmabling_time} for allowed values of $\Ll$. In bulk dimensions greater  than 3 we can choose the boundary subsystems in a way  which respects the axi-symmetry of the Kerr black hole, this may explain why the mutual information in the Kerr geometry sees a different dependence on the exponent of the blueshift $\kappa$.
\\\\
The physics of scrambling of mutual information in a TFD state is very similar to that of the decay of the OTOC in the same state \cite{Shenker:2013pqa}. In fact one can regard the mutual information to be a better estimate of the entanglement between the 2 CFTs of the TFD state than the correlation functions of local operators \cite{Wolf:2007tdq}. The study of SYK-like models suggests that late time scrambling is associated with a Schwarzian action at very low temperatures, this picture has also since been holographically realised in the near horizon dynamics of the JT model for  near extremal black holes \cite{Jensen:2016pah,Maldacena:2016hyu}. However the rate of scrambling in the JT model is controlled by the small infinitesimal temperature of the near extremal geometry. 
The above result suggests that for near extremal Kerr black holes there has to exist a similar IR description in terms of an effective 2d theory explaining scrambling at a rate greater than the temperature of the black hole. Note, this is has to be true for any allowed value $\Ll$.
This is to be expected as the JT model studied till now in the context of higher($d>2$) dimensional black holes describe the $J=const$ \footnote{$Q=const$ for RN \cite{Nayak:2018qej} and  $J=const,\,Q=const$ for Kerr-Newman.} sector \cite{Moitra:2019bub,Moitra:2018jqs,Castro:2018ffi,Castro:2019crn,Castro:2021fhc} . However, the rotating shockwaves with angular momentum per unit energy $\Ll$ tend to explore a different sector as the  geometry is perturbed along $\delta J=\Ll\,\delta\mathcal{M}$. 
In other words, if one were to study the IR effective gravity theory for near horizon dynamics of near extremal Kerr or Kerr-Newman geometries for $\delta J-\Ll\,\delta \mathcal{M}=0$, the above results suggest that the effective theory must account for scrambling at a rate greater than the temperature of the black hole and controlled by $\Ll$. This is also in some sense suggestive from the first law of black hole mechanics which in such a case takes the form
\be
\delta\mathcal{M}-\mu\,\delta J=T_H\delta \mathcal{S}\implies \delta\mathcal{M}=\frac{T_H}{1-\mu\,\Ll}\delta\mathcal{S}.
\ee
Further in case of RN black holes as the blueshift for rotating and non-rotating shockwaves stay the same, the near horizon dynamics describing the $Q\neq const$ sector should see the scrambling governed by the temperature of the black hole. 
\\\\
The phenomena of pole-skipping also serves as a hallmark of maximally chaotic many body quantum systems. Here the poles in the frequency space of the retarded energy density 2pt functions are skipped at points which directly correspond to $\lambda_L$ and $v_B$- the butterfly velocity \cite{Blake:2018leo}. 
For the case of rotating BTZ  this was first analysed in \cite{Liu:2020yaf} where pole skipping points implied 2 possible Lyapunov indices corresponding to left and right temperatures of the dual CFT$_2$ as found in \cite{Poojary:2018eszz,Jahnke:2019gxr}. 
The case for Kerr-AdS$_4$ was recently analysed \cite{Blake:2021hjj} and the Lyapunov index and the butterfly velocity were inferred by analysing the ingoing solutions to metric perturbations along null directions at the horizon. Here the Lyapunov index was found to be $\lambda_L=2\pi T_H=\kappa_0$ $i.e.$ the temperature of the black hole. 
This is consistent with our result as one expects to find the scrambling time governed by the temperature of the black hole $i.e.$ $\kappa_0$, for non-rotating shockwaves. 
The in-falling null coordinates used in \cite{Blake:2021hjj} are precisely non-rotating and- according to  the general arguments presented in section 2, must see a blueshift given by $2\pi T_H$. Our results further suggest that adapting the analysis of \cite{Blake:2021hjj} to rotating coordinates provided in section 2 one must be able see pole-skipping at points $\omega_*=i\kappa=2\pi i T_H/(1-\mu\,\Ll)$\footnote{The work in \cite{Blake:2021hjj} also tries to check the OTOC behaviour by analysing the growth of perturbations at the horizon due to a non-rotating shockwave.}. This amounts to studying the energy density or the stress-tensor response for in-falling perturbations with angular momentum per unit energy $\Ll$ at the horizon. It would be worth checking this explicitly as one also extracts the butterfly velocity from such an analysis.
\section*{Acknowledgement}  
RP would like to thank Daniel Grumiller and Prashanth Kocherlakota for discussions related to certain aspects of the project.
RP is supported by the Lise Meitner project FWF M-2882 N.
\appendix
\section{RN-AdS$_4$}
\label{RN}
Here we simply note the index of the blueshift suffered by a null rotating shockwave in Reissner-Nordstr\"{o}m black hole in $AdS_4$. 
The metric for RN-$AdS_4$ in Boyer Lindquist coordinates takes the form
\bea
ds^2_{RN}&=&\frac{dr^2}{f(r)^2}-dt^2f(r)^2+r^2(d\theta^2+\sin^2\theta \,d\phi^2)^2\cr&&\cr
{\rm where}\,\,\,
f(r)&=& 1-\frac{2M}{r}+\frac{4\pi Q^2}{r^2}+r^2,\hspace{0.4cm} F_{rt}=\frac{Q}{r^2}
\label{RN_AdS_4}
\eea  
RN-AdS$_4$ black holes with magnetic charges too have been considered \cite{Nayak:2018qej}, however the charge and $U(1)$ dynamics would play no role in deciding the blueshift at very late times.
The solutions for null geodesic are likewise obtained by solving the conditions \eqref{geodesic_cond} on the vector fields $\xi\cdot \partial$. As the geometry is spherically symmetric the analysis is easier and there is no Carter's constant to be specified, thus implying we can set $\xi^\theta$ component to be zero.
The in-out going null pairs are similarly constructed by reversing the signs of $\mathcal{E}$ \& $\Ll$. Taking $\mathcal{E}=1$ and writing the metric line element in terms of the duals to such null pairs of vector fields we have
\bea
ds^2_{RN}&=&F(r)dudv+h(r)\sin^2\theta(d\phi+h_1(r)d\tau)^2+r^2d\theta^2\cr&&\cr
{\rm where}\hspace{0.2cm}\tau&=&t-\Ll\,\phi,\cr&&\cr
F(r)&=&\frac{r^2 \left(r \left(-2 M+r^3+r\right)+4 \pi  Q^2\right)}{r^4-\mathcal{L}^2 \csc ^2 \theta  \left(r \left(-2 M+r^3+r\right)+4 \pi  Q^2\right)},\cr&&\cr
h(r)&=&r^2 \sin ^2 \theta -\frac{\mathcal{L}^2 \left(r \left(-2 M+r^3+r\right)+4 \pi  Q^2\right)}{r^2},\cr&&\cr
h_1(r)&=&\frac{\mathcal{L} \csc ^2 \theta  \left(r \left(-2 M+r^3+r\right)+4 \pi  Q^2\right)}{r^4-\mathcal{L}^2 \csc ^2 \theta  \left(r \left(-2 M+r^3+r\right)+4 \pi  Q^2\right)}
\eea
Note that here there is no redefinition of the $\phi$ coordinate required as $h_1$ goes to zero at the horizon. 
$\kappa$ is then similarly defined as in \eqref{Kappa} by demanding smoothness (affines) of the parameter along the null geodesics at the outer horizon 
\be
\kappa=\left.\mathcal{K}\right\vert_{\rpl}=\left|\tfrac{1}{2}\xi_\pm.\partial F\right\vert_{\rpl}=\frac{2\pi}{\beta}
\ee
Thus we see that rotating null geodesics see the same blueshift at the horizon as  non-rotating ones $i.e.$ the exponent of the blueshift $\kappa$ is the temperature of the RN-$AdS_4$ irrespective of the geodesics angular momentum.
 
\bibliographystyle{JHEP.bst}
\bibliography{bulk_syk_soft_modes.bib}
\end{document}